\documentclass[12pt]{article}

\usepackage[dvips]{color}
\usepackage{amsmath}
\usepackage{graphicx} 
\usepackage[dvips]{color}
\usepackage{graphicx}
\usepackage{tabularx}
\usepackage{subfig}
\usepackage[outercaption]{sidecap}    
\usepackage{hyperref}

\newcommand*{\linkme}[2]{\href{#1}{#2}\footnote{See \url{#1}}}

\begin{document}
\title{Quantum centipedes with strong global constraint}
\author{Pascal Grange\\
{\emph{Department of Mathematical Sciences}}\\
 {\emph{Xi'an Jiaotong-Liverpool University}}\\
{\emph{111 Ren'ai Rd, 215123 Suzhou, China}}\\
\normalsize{{\ttfamily{pascal.grange@xjtlu.edu.cn}}}}

\date{}
\maketitle
\vspace{1cm}
\begin{abstract}
 A centipede made of $N$ quantum
walkers on a one-dimensional lattice is considered. The distance between two consecutive legs is either
one or two lattice spacings, and a global constraint is imposed: the maximal distance between the first and last leg is $N+1$. This is the strongest global constraint compatible with walking. 
 For an initial value of the wave function 
 corresponding to a localized configuration at the origin, the probability law of the first leg of the 
 centipede can be expressed in closed form in terms of Bessel functions. The dispersion relation and the group velocities
 are worked out exactly. Their maximal group velocity goes to zero when $N$ goes to infinity, which is in contrast with the behaviour of group velocities of quantum centipedes without global constraint, which were
 recently shown by Krapivsky, Luck and Mallick to give rise to ballistic spreading 
 of extremal wave-front at non-zero velocity in the large-$N$ limit.  The corresponding Hamiltonians are implemented
 numerically, based on a block structure of the space of configurations corresponding to compositions 
 of the integer $N$. The growth 
 of the maximal group velocity when the strong constraint is gradually relaxed is
 explored, and  observed  to be linear in the density of gaps allowed in the configurations. 
  Heuristic arguments are presented to infer that the large-$N$ 
 limit of the globally constrained model can yield finite group velocities provided the allowed number 
 of gaps is a finite fraction of $N$. 
\end{abstract}
\tableofcontents

\section{Introduction and background}
Multi-pedal synthetic molecular systems made of single-stranded DNA segments
  moving on surfaces or tracks containing complementary DNA segments (see \cite{macromoleculesReview} for a review)
 have triggered the study of statistical-mechanical models of {\emph{molecular spiders}} or {\emph{centipedes}}
  whose  configurations are those of  a system with  $N$ {\emph{legs}} that 
 are all  attached to vertices of a regular lattice, but can step to the nearest sites under certain constraints, which can be
 local or global. On a one-dimensional lattice, the simplest local constraint forbids the attachment of more than one leg to the same site
 at the same time, and puts an upper bound $s$ on the distance between two adjacent legs. On the other hand, global constraints 
 forbid configurations in which the distance between the first and last legs are larger than some upper bound $S$. If the 
 spacing of the one-dimensional lattice serves as the unit distance, the simplest local constraint that still allows the centipede 
 to move is $s=2$. The classical dynamics of a centipede with this local constraint has been studied  in \cite{classicalCentipedes}, where the diffusion
 coefficient of the centipede is worked out as a function of the number of legs.\\


Quantum walks, on the other hand \cite{quantumWalks,quantumWalksReview,bifractality,twoBodyQuantum}, yield ballistic rather than diffusive behaviour.  Group velocities of quantum centipedes have been worked out 
 recently  by Krapivsky, Luck and Mallick \cite{quantumCentipedes}, with local constraints only, for all values of $N$, by mapping the quantum centipede to an integrable spin chain with boundaries. The distance between consecutive legs are 
 $N-1$  binary variables due to the local constraint.  Jordan--Wigner and Bogoliubov
 transformations applied to each Fourier mode of the wave function of the centipede have yielded eigenvalues of the Hamiltonian of Fourier modes in terms of numbers of quasi-particles. The maximal group velocity has been shown to reach a non-zero saturation 
 value at large $N$. In particular the maximal group velocity of the quantum centipede has been found to converge to a non-zero value in the large-$N$ limit. In this 
 paper, I consider the other extreme regime of global constraint by imposing the strongest global constraint ($S = N$) that still allows the centipede to walk.
 This gives rise to an exactly solvable model. When this constraint is gradually relaxed by allowing more
 gaps in the configurations of the centipede, the maximal group velocity allowed by the spectrum of the Hamiltonian 
 is expected to grow, interpolating between the results of the model with the strongest constraint and the saturation value.
 This growth is investigated numerically.\\



\section{The Schr\"odinger equation for a centipede with strong global constraint}
\subsection{Definitions and notations}
Consider a centipede with $N$ legs, moving along an infinite one-dimensional oriented
regular lattice with unit spacing, and vertices labelled by integers, such that:\\
- each leg is attached to a vertex of the lattice, and at most one leg is attached to any given site (attachment constraint),\\
- any  two consecutive legs are separated by at most one empty 
vertex, or {\emph{gap}} (local constraint).\\
For a fixed position of the first (i.e. leftmost) leg, a configuration can be depicted by a sequence
of regularly-spaced black and white bullets, with each black bullet corresponding to a leg attached to a vertex, and each white bullet
corresponding to an empty vertex. Because of the attachment constraint, an allowed configuration contains
exactly $N$ black bullets. For example, in the case $N=3$, the configuration of the legs allowed by the attachment constraint and the local constraint are the following:
\begin{equation}
(\bullet\bullet\bullet),\; (\bullet\circ\bullet\bullet),\; (\bullet\bullet\circ\bullet),\; (\bullet\circ\bullet\circ\bullet),
\label{config3}
\end{equation}
where the leftmost black dot on the left of each configuration is the first leg of the centipede (there are infinitely 
 many empty vertices on the left of every configuration, as the lattice is infinite in both directions).\\

 We can add a global constraint by requiring that the two external 
legs are at a distance at most $S$ from each other. The total length of the centipede is therefore
 between $N-1$ and $S$. The bound $S$ cannot be lower than 
$N-1$ because of the attachment constraint. If $S=N-1$, there is only one configuration allowed
once the position of the first leg is chosen. Hence the value $S = N$ is the strongest global 
 constraint on the length of the centipede still allowing the centipede to move.\\ 
 In this note we focus on this case ($S=N$).  Let us describe leg configurations.
 The only parameter in the model is 
 $N$. If $N=2$, the local and global constraint are equivalent,
 and we are left with the two-body problem studied in \cite{twoBodyQuantum}. Let 
 us therefore assume that $N\geq 3$. 
  There are $N$ allowed leg configurations for the centipede (if the position of the first leg is fixed):\\
- one in which there is no empty vertex between any pair of legs (the total length of the centipede is 
 $N-1$ in this configuration), let us call it $l_0$;\\
- configurations in which exactly one pair of 
consecutive legs, say the $k$-th and $(k+1)$-th legs, are separated by one empty vertex;
 let us call such a configuration $l_k$, where $k$ can take all integer values between $1$ and $N-1$.\\
In the special case $N=3$, the first three configurations depicted in Eq. \ref{config3} are still
 allowed by the global constraint (they correspond to $l_0$, $l_1$ and $l_2$ respectively), while the last one is forbidden.\\


\subsection{Schr\"odinger equation for the Fourier modes of the quantum centipede}

 The allowed moves in an infinitesimal time step are of the following kinds.\\

 $(i)$ From the configuration $l_0$ to either configuration $l_1$ or configuration $l_{N-1}$ (in the
 case $N=3$, these moves map $(\bullet\bullet\bullet)$ to $(\bullet\circ\bullet\bullet)$ and $(\bullet\bullet\circ\bullet)$ respectively), raising the length of the centipede  to $N$, the largest value allowed by the global constraint.\\

$(ii)$ From the configuration $l_k$ (with $k$ an integer in $[1..N-2]$ to either
 the configuration $l_{k-1}$ or the configuration labelled $l_{k+1}$  (in the
 case $N=3$, these moves map $(\bullet\circ\bullet\bullet)$ to  $(\bullet\bullet\bullet)$  and 
$(\bullet\bullet\circ\bullet)$ respectively, either conserving the length of the centipede, or reducing ot by one).\\

 $(iii)$ From the configuration  $l_{N-1}$ to either the configuration  $l_{N-2}$
 or to the configuration $l_{0}$ (in the
 case $N=3$, these moves map $(\bullet\bullet\circ\bullet)$ to  $(\bullet\circ\bullet\bullet)$ and $(\bullet\bullet\bullet)$ respectively, conserving the
 length of the centipede in the first case, reducing the length  by one in the second case, but without moving the first leg, unlike the  move described in $(ii)$ that maps the configuration to $l_0$).\\

 Let us describe the quantum walk of the centipede by the wave function $|\psi(t)\rangle$, which 
 lives in the Hilbert space $\mathcal{H}_N$ consisting of the tensor product of the position of the first leg and the
 space spanned by the above-described configurations of legs:\\
\begin{equation}
|\psi(t)\rangle \in \mathcal{H}_N = \mathbf{Z}\otimes {\mathrm{Span}}_{\mathbf{C}}( \left\{ |l_k\rangle \right\}_{0\leq k \leq N-1}).
\end{equation}
where $|l_k\rangle$ is the ket corresponding to the leg configuration denoted by $l_k$.\\

If the position of the first leg is fixed at the vertex labelled $n$ (corresponding to the ket $|n\rangle$), the wave function 
 of the centipede at time $t$ can therefore be described as an element of the projection of the Hilbert space $\mathcal{H}_N$
 onto the ket $|n\rangle$:\\
\begin{equation}
 |\psi_n( t )\rangle = \langle n |\psi(t)\rangle.
\end{equation}
 Let us introduce the Fourier transform of the wave function, by defining for every wave
 number $q$ in the first Brillouin zone
\begin{equation}
| \hat{\psi}(q, t ) \rangle= \sum_{n \in \mathbf{Z}}  e^{-inq}|\psi_n(t)\rangle.
 \label{Fourier}
\end{equation}
 The ket $| \hat{\psi}(q, t ) \rangle$ is an element of the $N$-dimensional span of 
 the leg configurations:\\
\begin{equation}
| \hat{\psi}(q, t ) \rangle \in  {\mathrm{Span}}_{\mathbf{C}}( \left\{ |l_k\rangle \right\}_{0\leq k \leq N-1}).
\end{equation}

For a fixed wave number $q$, the allowed moves in continuous time (with identical 
 rates) that have been described in $(i)$--$(iii)$ induce a Schr\"odinger equation for 
 the Fourier transform $|\hat{\psi}(q, . )\rangle$, with a $q$-dependent Hamiltonian:
\begin{equation}
 i \frac{\partial}{\partial t}| \hat{\psi}(q, t ) \rangle = H(q)  | \hat{\psi}(q, t ) \rangle,
\label{Schrodinger}
\end{equation}
 where the Hamiltonian operator $H(q)$ is  expressed as
\begin{equation}
 H(q) = J(q) + J(q)^\dagger,
\end{equation}
with  the entry of the $N$ by $N$ matrix of $J(q)$ at row $i$ and column $j$ (in the basis $\left\{ |l_k\rangle \right\}_{0\leq k \leq N-1}$) is given by
\begin{equation}
(J( q ))_{ij} =   e^{-iq}\delta_{i1}\delta_{j2}+ \delta_{iN}\delta_{j1}  + \sum_{j=3}^N\delta_{i,j-1}.
\end{equation}
In the case $N=3$, we obtain\footnote{As $N$ is the only parameter in the model,
 we exclude the symbol $N$ from most notations, making the dependence on $N$ implicit, but restore it for definiteness when describing the case $N=3$.} the following matrices:
\begin{equation}
J(q)_{(N=3)} =  \left(
   \begin{array}{lll}
       0 &  e^{-iq}& 0 \\
      0  & 0 & 1\\
      1&  0  & 0 
   \end{array}
   \right),\;\;\;
H(q)_{(N=3)} =  \left(
   \begin{array}{lll}
       0 &  e^{-iq}& 1\\
      e^{iq}  & 0 & 1\\
      1&  1  & 0 
   \end{array}
   \right).
\end{equation}
The $q$-dependent matrix elements in the Hamiltonian reflect the fact that the only allowed moves that modify the position of the first leg of the centipede
 are  the moves from configuration $l_0$ to configuration $l_1$, and from $l_1$ to configuration $l_0$ (together with the 
 definition of the Fourier transform, Eq. \ref{Fourier}). The rules assigning the values $e^{iq}$, $e^{-iq}$
 or $1$ to matrix elements of the Hamiltonian 
 according to configuration moves 
 are identical to the ones worked out in \cite{quantumCentipedes} (see Eqs. (3.7) in Section 3 of that paper). 
 The effect of the global constraint 
 is just to reduce the set of allowed configurations and moves, ensuring that the  Hamiltonian  $H(q)$ 
 is a submatrix extracted from the matrix of size $2^{N-1}$ corresponding to the  dynamics of
  a centipede with the same number of quantum walkers, identical local constraints and no global constraint.\\

\subsection{Group velocities}
 The Hamiltonian $H(q)$ can be diagonalised using the same method as for 
 a circulant matrix (to which it reduces if $e^{iq}$ equals 1). Indeed, we 
 can diagonalise $J(q)$ and $J(q)^\dagger$ separately, and notice that 
 they have the same eigenvectors (associated to complex conjugated 
 eigenvalues).  These eigenvectors are therefore those of the Hamiltonian,
 and they are associated to eigenvalues equal to twice the real part of those of $J(q)$.\\

Let us denote by $|u\rangle =\sum_{j=0}^{N-1} u_j |l_j\rangle$ an eigenvector of $J(q)$, and by $\lambda$ the associated eigenvalue.
The components  satisfy the system of equations:
$$ e^{-iq} u_1 = \lambda u_0,$$
$$ \forall j \in [ 1..(N-2)],  \;u_{j+1} = \lambda u_j ,$$
$$ u_0 = \lambda u_{N-1}.$$
This system implies 
$$ \forall j \in [ 1..(N-1)],\; u_j = \lambda^{j-1} e^{iq} u_0,$$
$$ u_0 = \lambda^N e^{iq} u_0.$$
 The component $u_1$ cannot be zero, otherwise the eigenvector $|u\rangle$ would be zero,
 hence the eigenvalue $\lambda$ must satisfy
\begin{equation}
 \lambda^N e^{iq} = 1.
\end{equation}
The eigenvalues of $J(q)$ are therefore elements of the set  of complex numbers
\begin{equation}
 \left\{\lambda_k (q)= e^{\frac{i}{N}\left(-q + 2k\pi\right)}, \; k \in [ 0.. N-1] \right\}.
\label{eigenVals}
\end{equation}
 Moreover, using the  system of equations, one can check that for fixed $k$ in $[ 0.. N-1]$  the ket  
\begin{equation}
|u^{(k)}(q)\rangle = \frac{1}{\sqrt{N}} \sum_{j=0}^{N-1}\lambda_k(q)^{j}e^{-iq\delta_{j0} }|l_j \rangle
\label{eigenVects}
\end{equation}
 is a normalized eigenvector of $J(q)$ associated with the eigenvalue $\lambda_k(q)$. Moreover, it is an eigenvector
 of $J(q)^\dagger$ associated with the eigenvalue $\lambda_k^{-1}$. It is therefore an eigenvector 
 of the Hamiltonian $H(q)$ associated with the eigenvalue $\lambda_k + \lambda_k^{-1}$. The spectrum of $H(q)$
 therefore consists of $N$ eigenvalues expressed as cosines with regularly spaced arguments:
\begin{equation}\label{dispersion}
{\mathrm{Spec}}\left(H(q)\right) = \left\{  \omega_k(q) = 2\cos\left( \frac{1}{N}\left( - q + 2k\pi\right)\right),\;k\in [0..N-1] \right\}
\end{equation}
The group velocities corresponding to these modes read
\begin{equation}
v_k(q) = \frac{\partial}{\partial q}\omega_k(q) = \frac{2}{N} \sin\left( \frac{1}{N}\left( - q + 2k\pi\right) \right), \;k\in [0..N-1].
\end{equation}
The maximal absolute value $V^{(N)}$ of the group velocities of the strongly 
 constrained $N$-legged quantum centipede
\begin{equation}
 V^{(N)} = \mathrm{max}_{q\in [-\pi, \pi], k \in [ 0..N-1 ]} |v_k(q)|
\end{equation}
 therefore goes to zero in the large-$N$ limit. If $N$ is a multiple of 4, the sine function 
reaches its maximal value at $q=0$, along the branch corresponding to $k = N/4$. If $N$ 
is large enough, considering the division of $N$ by 4 is enough to see that the maximal
 possible value of the group velocities is always reached.

\section{Solution of the Schr\"odinger equation for a centipede  with strong global constraint starting from the origin}
\subsection{Expression of the wave fuction in a basis adapted to Fourier modes}
Consider the initial condition corresponding to a centipede with the first leg at the origin and 
 no empty vertex between any pair of legs:
\begin{equation}
|\psi_n(0) \rangle= \delta_{n,0} |l_0\rangle,\;\forall n \in {\mathbf{Z}}.
\end{equation}
Rewriting the position of the leftmost leg as an integral over Fourier 
modes, and using the basis $\left\{ |l_k\rangle \right\}_{0\leq k \leq N-1}$ of the space of leg configurations, we can integrate the Schr\"odinger equation (Eq. \ref{Schrodinger}) as follows:
\begin{equation}
|\psi_n(0) \rangle= \int_0^{2\pi} \frac{dq}{2\pi} e^{inq} \sum_{k=0}^{N-1}\langle u^{(k)} |l_0\rangle |u^{(k)}(q)\rangle,
\end{equation}
\begin{equation}
|\psi_n(t) \rangle= \int_0^{2\pi} \frac{dq}{2\pi} e^{inq}\sum_{k=0}^{N-1}\langle u^{(k)}(q) |l_0\rangle  e^{-i\omega_k t} |u^{(k)}(q)\rangle.
\end{equation}
Substituting the results of the diagonalisation of $H(q)$, Eqs \ref{eigenVals} and \ref{eigenVects}, we express the ket
 $|\psi_n(t)\rangle$ as a  sum over eigenvectors:
\begin{equation}
|\psi_n(t) \rangle= \frac{1}{\sqrt{N}}\int_0^{2\pi} \frac{dq}{2\pi} e^{i(n+1)q}\sum_{k=0}^{N-1}  e^{-2it\cos\left( \frac{-q+ 2k\pi}{N}\right)} |u^{(k)}(q)\rangle.
\end{equation}

\subsection{Probability law of the first leg of the centipede}
The probability law of the position of the first leg at time $t$ is obtained  by adding
together the probabilities of the configurations with the same position of the first leg:
\begin{equation}
P_n(t) = \sum_{p=0}^{N-1} |\langle l_p | \psi_n(t) \rangle|^2 = \frac{1}{N}\sum_{p=0}^{N-1}\left| \int_0^{2\pi}
\frac{dq}{2\pi}\sum_{k=0}^{N-1} e^{i(n+1)q-2it\cos\left( \frac{-q+ 2k\pi}{N}\right)} \langle l_p|u^{(k)}(q)\rangle \right|^2.
\end{equation}
Using Eq.  \ref{eigenVects} for the components of the eigenvectors of $H(q)$ yields integrands in exponential 
form:
\begin{equation}
P_n(t) =  \frac{1}{N^2}\sum_{p=0}^{N-1}\left| \int_0^{2\pi}
\sum_{k=0}^{N-1}\frac{dq}{2\pi} e^{i(n+1)q-2it\cos\left(\frac{-q+ 2k\pi}{N} \right)} \lambda_k(q)^{p}
(e^{-iq}\delta_{0p} + (1-\delta_{0p})) \right|^2.
\end{equation}
In the integral corresponding to index $k$, let us perform the change of variable
\begin{equation}
K =\frac{-q+2k\pi}{N},
\end{equation}
which results in $k$-dependent bounds in the integrals (with no dependence on the integer $k$ left in the integrand, due to
 periodicity):
\begin{equation}
P_n(t) = \sum_{p=0}^{N-1}\left| \sum_{k=0}^{N-1} \int_{\frac{2k\pi}{N}}^{\frac{2(k+1)\pi}{N}}
\frac{dK}{2\pi} e^{i(-(n+1)N+p)K-2t\cos  K )} (e^{iNK}\delta_{0p} + (1-\delta_{0p})) \right|^2.
\label{PnDouble}
\end{equation}
The sum over the integer $k$ can therefore be performed by extending the integration
 domain to the full interval $[0,2\pi]$,  yielding Bessel functions according to the 
  integral definition:\\
\begin{equation}
J_m(t) = i^m\int_0^{2\pi} \frac{dq}{2\pi} e^{i(mq - t \cos q )},
\end{equation}
so that each value of the index $p$ in Eq. \ref{PnDouble} yields a  Bessel function:
\begin{equation}
P_n(t) = \left(|J_{(-nN)}(2t)|^2 + \sum_{p=1}^{N-1}| J_{p-(n+1)N }(2t)|^2\right).
\end{equation}
This closed-form expression is reminiscent of the case of a single quantum walker, in which 
 only one Bessel function appears in the wave function \cite{twoBodyQuantum}.\\


\section{Growth of the maximal group velocity as a function of the maximal density of gaps}
\subsection{Mapping  configurations of  constrained centipedes to compositions of the number of legs}

 Consider  a centipede with $N$ legs, and allow a weaker global constraint on the maximal 
 total distance between the first and last legs, $S\geq N$. Let us keep the local constraint 
  on the  maximal number of empty sites between two consecutive legs.
 The  maximal total number $G$ of gaps allowed 
 between legs is $G = S-N+1$.  The case $G=1$ has been solved exactly, let us now assume 
 $G>1$. A  basis of the Hilbert space is given by the direct sum 
 of spaces of linear combinations of configurations with a fixed number $g$ of gaps, for $g$ in $[0..G]$. The value $g=0$ 
 corresponds to just one configuration, with length $N-1$. At a fixed
 value $g>1$, the allowed configurations are in one-to-one correspondence
 with the number of ways of  choosing $g$ distinct integers $p_1,\dots,p_g$, with 
 $1\leq p_1 < \dots < p_g < N$,
 where $p_k$ is the index of the $k$-th leg that is followed by a gap. If $g>1$, we can rewrite the last index $p_g$ 
 by grouping consecutive indices as follows:\\
\begin{equation}
p_g = p_1 + \sum_{k=2}^g( p_k - p_{k-1}),
\end{equation}
which induces a $(g+1)$-composition\footnote{A composition of an integer $n$ with $k$ terms (or $k$-composition) is a 
 way to write $n$ as a sum of $k$ strictly positive integers, i.e. an element 
 $(u_1,\dots, u_k)$ of $({\mathbf{N}}^\ast) ^k$, such that $n = \sum_{p=1}^k u_p$. 
  The  order of the terms is taken into account: two $k$-compositions $(u_1,\dots, u_k)$ and $(u'_1,\dots, u'_k)$
 of $n$ are regarded as identical if and only if $\sum_{p=1}^k|u_p - u'_p| = 0$. If $n>1$ and $k>1$, the 
 $k$-compositions of $n$ are in one-to-one correspondence with a choice of $k-1$ spacings between 
 consecutive integers in $[1..n]$ (in our case, the role of $k-1$ is played by the number of gaps). 
 As there are $n-1$ such spacings, there are ${{n-1}\choose{k-1}}$ distinct $k$-compositions of $n$. This
 expression also yields the correct result  if $k=1$.} of $N$:
\begin{equation}
N = p_g + N - p_g = p_1 + \sum_{k=2}^g( p_k - p_{k-1}) + (N - p_g),
\end{equation}
corresponding to a description of the configuration in terms of the cardinalities  of sets of consecutive
 legs that do not contain any gap, read from left to right.\\

\begin{figure}
\begin{center}
\includegraphics[width= 18cm,keepaspectratio]{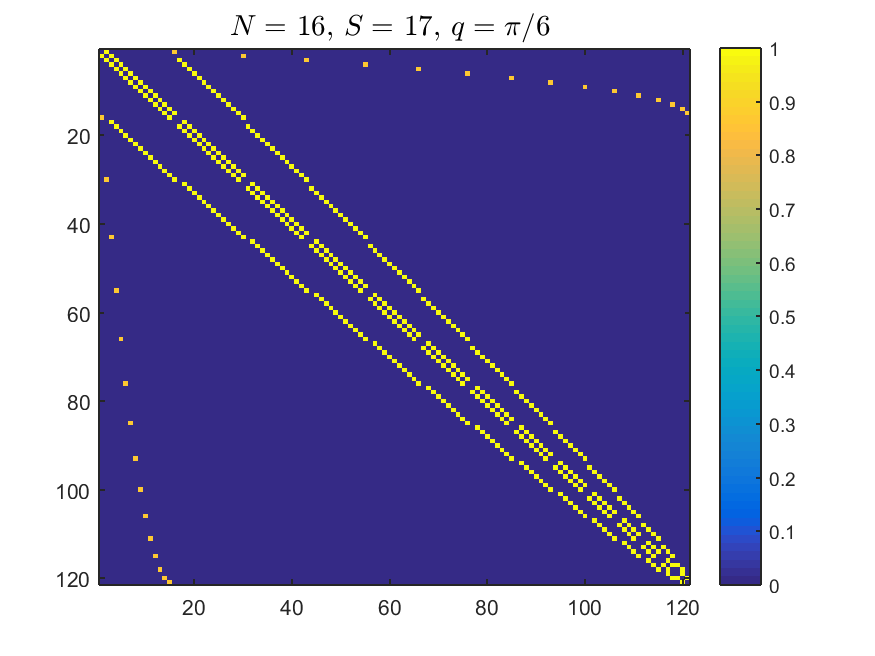}
\end{center}
\caption{ A heat map of the real part of the  Hamiltonian $H(q)$, for a centipede  with $N=16$ legs and a maximal of two gaps in any allowed configuration. The 
 chosen basis corresponds to diagonal blocks adapted to $k$-compositions of $16$ (arranged in lexicographic order), 
 for $k\in[1..3]$. The size of the matrix is
 $121 = 1 +  {{15}\choose{1}} + {{15}\choose{2}}$. The entries with absolute value equal to $\cos(q)$  and $\sin(q)$ correspond to couplings
 between configurations with different numbers of gaps.}
\label{figureHamiltonian}
\end{figure}

 Conversely, given a $(g+1)$-composition $(s_1,\dots,s_{g+1})$ of $N$, it can be mapped to an 
allowed configuration with $g$ gaps, by concatenating $(g+1)$ sets of legs with $s_k$ legs each, spaced
 by one gap. The allowed configurations at fixed length $N-1+g$ are therefore in one-to-one correspondence
 with the $(g+1)$-compositions of $N$. Let us denote by $\mathcal{C}_{g,N}$ the corresponding subspace
 of the Hilbert space. The Fourier modes of the wave function of the constrained quantum centipede 
  are therefore elements of the direct sum of these spaces, for all values of $g$ compatible with the 
 global constraint:
\begin{equation}\label{HilbertCompose}
|\hat{\psi}(q,t)\rangle \in \bigoplus_{g=0}^{S-N+1} \mathcal{C}_{g,N},
\end{equation}
 and from the above discusion
\begin{equation}\label{HilbertCompose}
{\mathrm{dim}}\;\mathcal{C}_{g,N} ={{N-1}\choose{g}}.
\end{equation}
The Hamiltonian $H(q)$ can be expressed in matrix form, with a block structure adapted to the decomposition 
 of the Hilbert space in Eq. \ref{HilbertCompose}. This requires a choice of order in 
 each of the relevant sets of compositions, but the spectrum 
 of the Hamiltonian does not depend on this choice. For numerical implementation
 we can choose the lexicographic order.
 The general rules allowing the centipede to move one leg at a time induce 
  couplings within a given diagonal block corresponding to compositions with a 
 fixed number of terms differing by the transfer of one unit between two adjacent terms.
 The corresponding entries in the matrix of the Hamiltonian equal one. All the non-zero elements  in 
 non-diagonal blocks  couple configurations belonging to blocks $\mathcal{C}_{g,N}$
 and $\mathcal{C}_{g',N}$, with $|g-g'|=1$.  Such couplings involve pairs of compositions,
 one of which has a $1$ in initial (resp. final) position (in which case the corresponding entry in the Hamiltonian
 equals $e^{\pm iq}$, resp. $1$) or terminal position.
 A heat map of the real part of the Hamiltonian with these 
 conventions for $N=16$ and $S=17$ (corresponding to a maximal of $G=2$ two gaps in configurations)
  is  illustrated  in Fig. \ref{figureHamiltonian}.\\

In the  notations introduced above in the case of the strongest constraint $S=N$, the integer 
 $j$ labelling the configurations  yields a basis of the one-dimensional space $\mathcal{C}_{0,M}$ 
 for $j=0$, while each of the remaining $N-1$ values of $j$ corresponds to the two-composition $N = j + (N-j)$.
  As the configuration with no gaps was put in the first position in the basis of the Hilbert 
  space in Section 2, the presentation of the Hamiltonian in matrix form 
 so far is consistent with the block structure adapted to compositions of $N$.\\

\subsection{Group velocities from perturbation theory}

\begin{figure}[!ht]
    \subfloat[\label{subfig-1:dummy}]{%
      \includegraphics[width=0.5\textwidth]{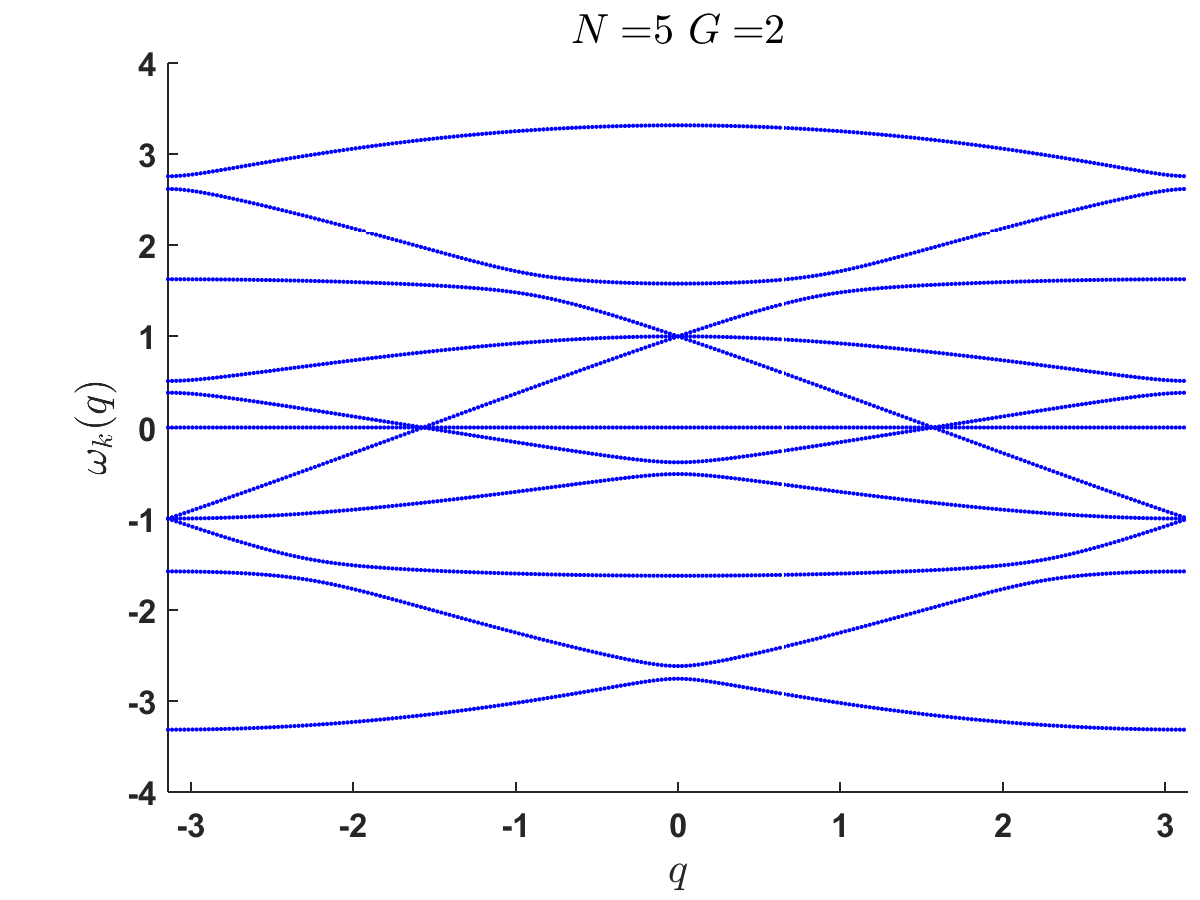}
    }
    \subfloat[\label{subfig-2:dummy}]{%
      \includegraphics[width=0.5\textwidth]{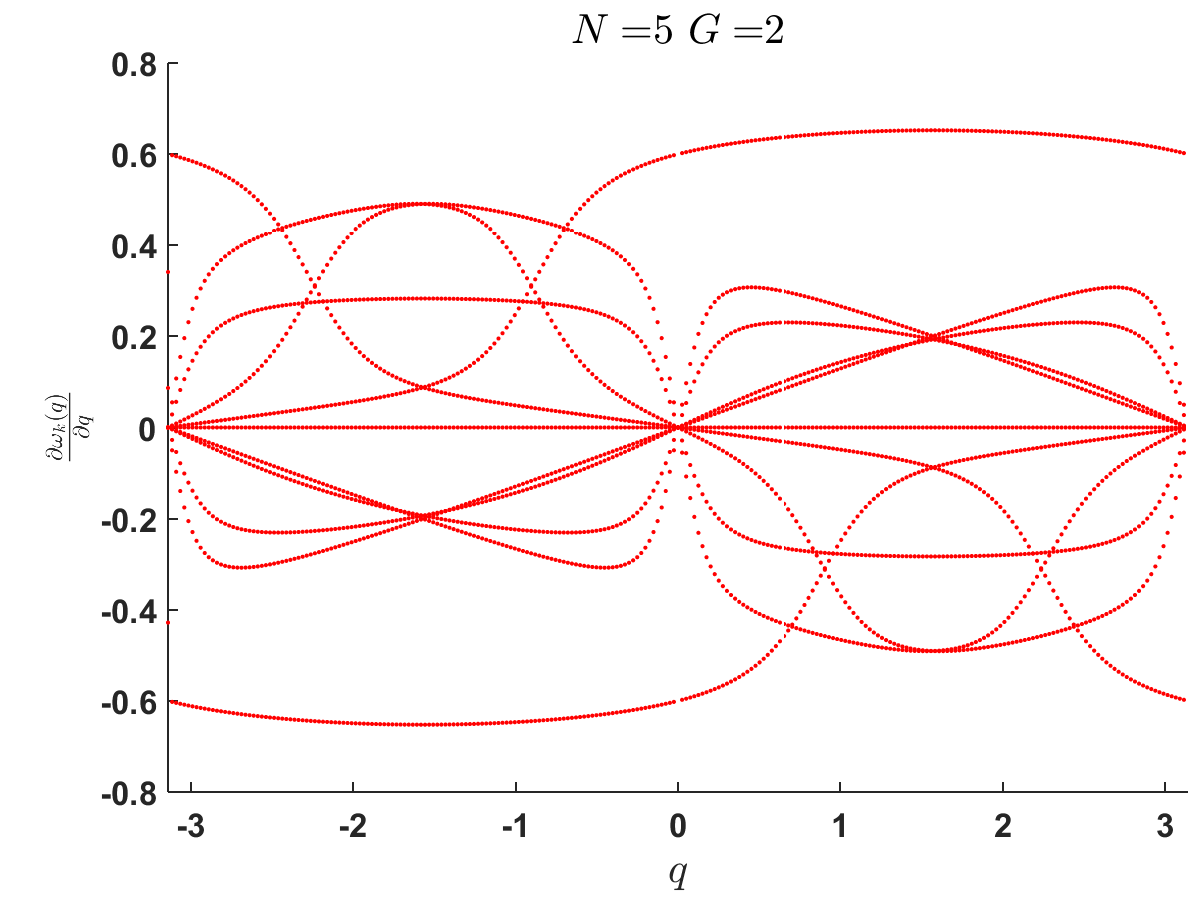}
    }
    \hfill
   \subfloat[\label{subfig-2:dummy}]{%
      \includegraphics[width=0.5\textwidth]{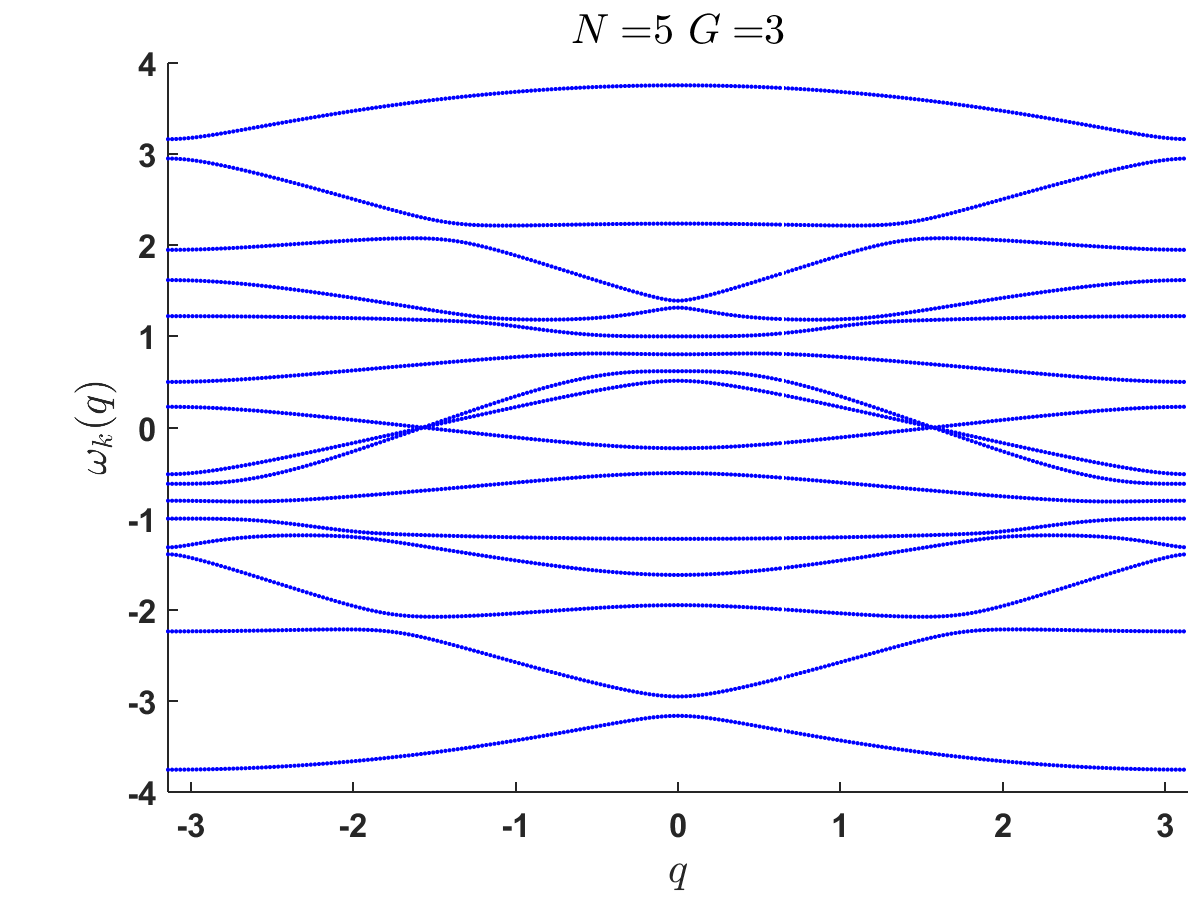}
    }
   \subfloat[\label{subfig-2:dummy}]{%
      \includegraphics[width=0.5\textwidth]{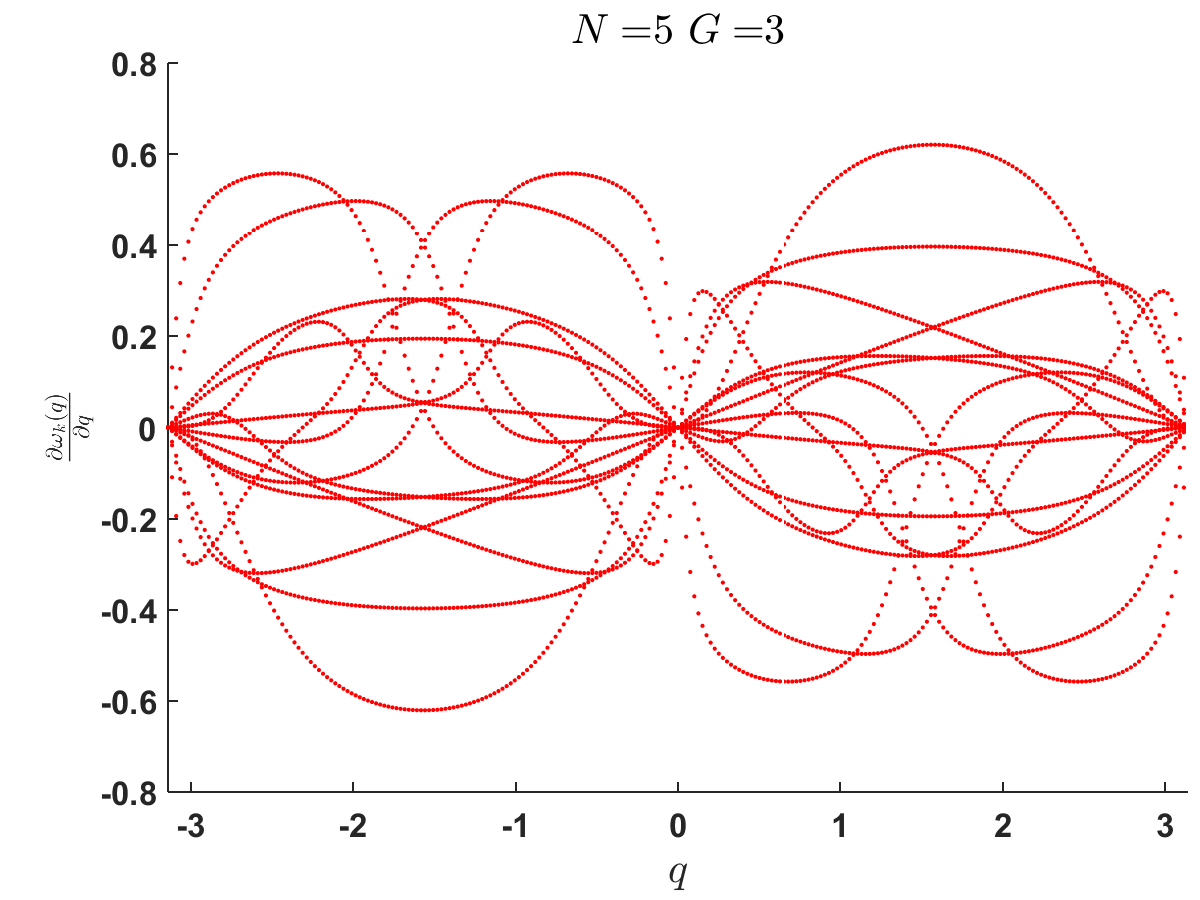}
    }
    \caption{ Phase velocities and group velocities for $N=5$ centipedes 
 corresponding to weaker global constraint that the exactly solved case.
 The group velocities have been computed at regularly spaces modes using first-order perturbation theory. There are $11$ 
  branches in the spectrum if up to $G=2$ gaps are allowed, and $15$ if up to $G=3$ gaps are allowed, in agreement with Eq. \ref{HilbertCompose}.}
    \label{figVelocities}
  \end{figure}

\begin{figure}
\begin{center}
\includegraphics[width= 16cm,keepaspectratio]{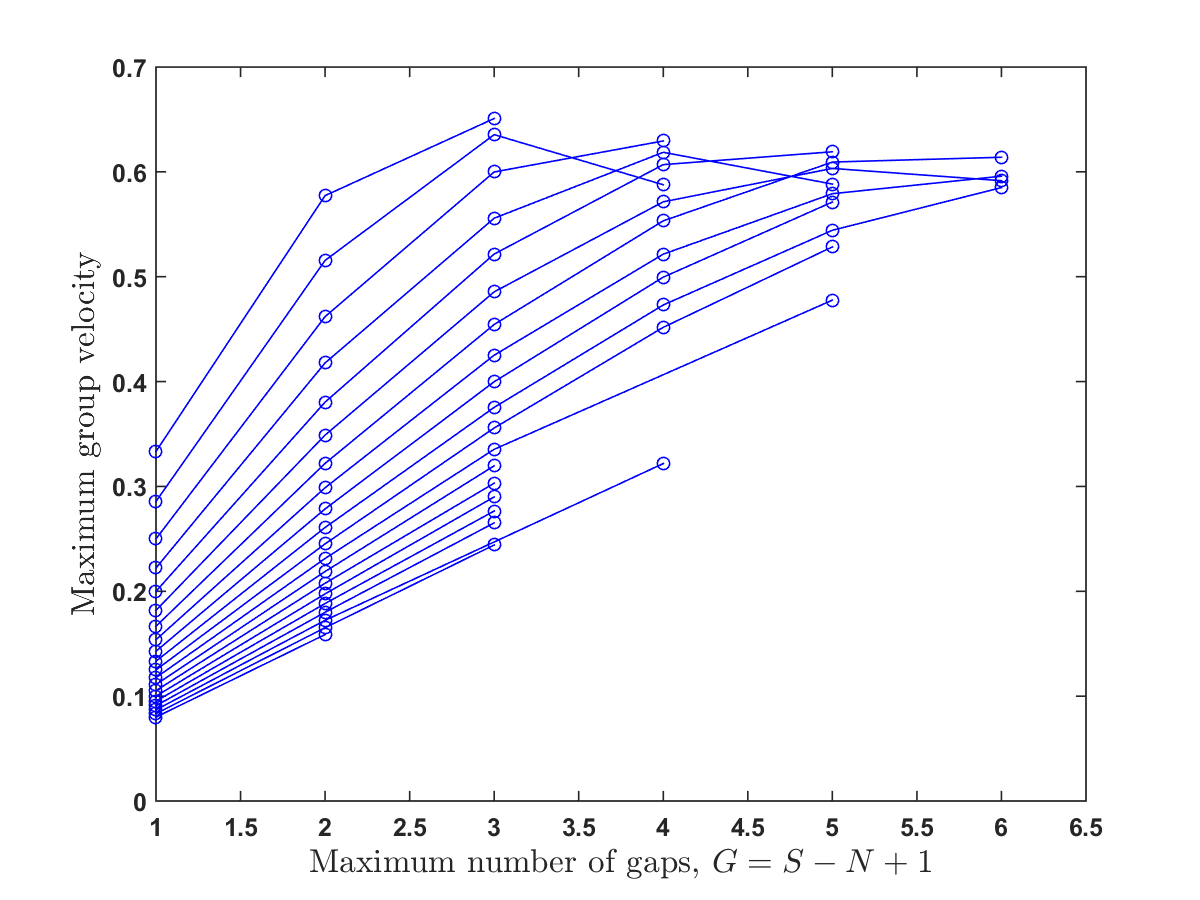}
\end{center}
\caption{The maximal group velocity as a function of the number of allowed 
 gaps in the configuration for $N$ between 6 and 25. A given curve is associated to a unique value of $N$,
 and the value at $G=1$ is $2/N$, so that curves associated to larger $N$ start at lower points. The tendency to affine growth 
at small values of the number of gaps is visible, as well as the saturation.}
\label{linearityCheck}
\end{figure}

 Given the number $N$ of legs  of the centipedes, and the total number $G$ of gaps between legs
 allowed by the global constraint, assuming a choice of order in the elements of the basis, let us denote by $H(q)$
  the  matrix form of the Hamiltonian. The $q$-dependent entries
  are  equal  either to $e^{iq}$ or to $e^{-iq}$.
 Let us denotes by $B$ 
 the sub-matrix of $H(q)$ containing only entries equal to 0 or 1 for all values of the mode $q$. 
 There is a matrix unique $A$ with entries equal to 0 or 1, such that the following relation holds for all
 $q$ in $[-\pi,\pi]$:
\begin{equation}
 H( q ) = e^{iq} A + e^{-iq}A^\dagger + B.
\end{equation}
 Given the spectrum of phase velocities and associated eigenvectors of $H(q)$, labelled by the integer $k$ 
 and a functional dependence on the mode $q$, the group velocities at mode $q' = q+\epsilon$ close to $q$  can be computed 
  using first-order perturbation theory (assuming non-degeneracy):
\begin{equation}
 H( q ) |\psi_{(k)}(q)\rangle = \omega_k(q) |\psi_{(k)}(q)\rangle, \;\;\forall q \in [-\pi, \pi],
\end{equation}
 \begin{equation}
 H( q + \epsilon ) = H(q) + \epsilon( ie^{iq} A - e^{-iq}A^\dagger) + o(\epsilon),
\end{equation}
so that the group velocities at mode $q$ are expressed as 
 \begin{equation}
 \frac{\partial \omega_k}{\partial q}(q)  = \langle \psi_{(k)}(q) | (ie^{iq} A - i e^{-iq}A^\dagger) | \psi_{(k)}(q)\rangle.
\end{equation}
The implementation\footnote{in MATLAB, code available online at Researchgate, {\ttfamily{https://www.researchgate.net/publication/315625145}}}, with regularly spaced modes in the
 interval $[-\pi,\pi]$, yields the spectra plotted in Fig. \ref{figVelocities} for $N=5$. 
 It allows to reproduce the relevant figures of \cite{quantumCentipedes} in the case $G=N-1$, corresponding 
 to a fully relaxed  global constraint. When the maximal group velocity is plotted at fixed value of $N$ as a function of $G$,
 the rate of growth can be observed between the value $2/N$ and the saturation value predicted 
  in \cite{quantumCentipedes}. The plot of Fig. \ref{linearityCheck} exhibits a linear growth at small 
 values of $G$. The propagation of wave packets yield heuristic arguments explaining this behaviour.

%
%


\subsection{Linear growth of the  group velocity with the density of gaps  and lower bounds on the large-$N$ limit}

Consider a quantum centipede with a large number $N$ of legs and a global constraint 
 corresponding to a maximal number $G$ of gaps. Let us assume that $G$  is a divisor 
 of $N$, so that there exists configurations with $G$ gaps obtained by concatenating $G$ identical configurations 
 of a centipede with $L=N/G$ legs and one gap. There must also be wave packets  with maxima
 spaced by $L$.  If $L$ is large enough, the maxima of these wave packets 
 should be well separated, and as the number of maxima equals the maximal number of
 gaps, the propagation of such wave packets should be well approximated at short times
 by the concatenation of $G$ wave packets, each of which is propagating 
 on a maximally constrained centipede with $L$ legs.\\
 
 Consider such a wave packet, described by a wave function $\phi$ in Fourier space, assumed to be peaked at the value $q_0$ of the linear momentum,
 and containing only modes from the branch of index $k$ of the spectrum 
 of the maximally constrained Hamiltonian,
 such that $\partial_q\omega_k(q_0)$ is the maximal group velocity.
  The corresponding wave function in position space, denoted by $\tilde{\phi}$, can be 
 expressed in terms of its projections on the kets $|l_p\rangle$, for $p$ between $0$ 
 and $L-1$:
 \begin{equation}
\begin{split}
 \langle l_p | \tilde{\phi}_n \rangle  (t) &= \int_0^{2\pi} dq \,\phi(q) e^{i(qn-\omega_k(q)t )}\langle l_p  |u_k(q)\rangle\\
& = \frac{1}{\sqrt{L}}\int_0^{2\pi} dq\,  \phi(q) \exp\left(i\left(q(n + \delta_{j0}-\omega_k(q)t +\frac{p}{L}(q+2k\pi)\right)\right).
 \end{split}
 \end{equation}
 With the dispersion relation of Eq. \ref{dispersion}, the derivatives of all even orders of $\omega_k(q)$
 vanish at the extremum $q_0$, while the derivatives of odd order $(2j+1)$  are proportional to the 
 maximal group velocity:
\begin{equation}
 \frac{\partial^{2j+1}\omega}{\partial x^{2j+1}}(q_0) = \frac{2}{L^{2j+1}}.
\end{equation}
The time-evolution of a wave packet traveling at the maximal group velocity
 is  therefore given by 
\begin{equation}
\begin{split}
 \langle l_p | \tilde{\phi}_n \rangle  &= \int_0^{2\pi} dq \,\phi(q) e^{i(qn-\omega_k(q)t )}\langle l_p  |u_k(q)\rangle\\
& = \frac{1}{\sqrt{L}}\int_0^{2\pi} dq\,  \phi(q) \exp\left(i\left(q(n + \delta_{p0}) -  2\sinh\left( \frac{ q-q_0 }{L} \right)t +\frac{p}{L}(q+2k\pi)\right)\right).
 \end{split}
 \end{equation}
 If the density of modes $\phi$ is sufficiently peaked around $q_0$, the wave front 
travels at velocity $2/L$.   As the space is discrete, the characteristic time step 
 for this evolution to be detectable is $\delta t = L/2$.
 For the concatenation of   $G$ such short wave packets to yield a wave packet
 with fronts traveling at velocity $2/L$, the deformation of the wave packet $\tilde{\phi}$ 
 must be small over the time interval $\delta t$. A large-$L$ expansion at first non-trivial order yields
\begin{equation}
 \langle l_p | \tilde{\phi}_n \rangle (\delta t)  = \left( e^{2i\frac{q_0}{\delta t}} \langle l_p | \tilde{\phi}_{n-\frac{2\delta t}{L}} \rangle (0) \right) \left(  1 + O( L^{-3}\delta t\right) ),
\end{equation}
  which shows that the description of the propagation as a  traveling wave packet, over a time step $\delta t = L/2$, should become better
 for larger $N$ at fixed $G$.\\ 
%
 

 In real space, starting from a concatenation of $G$ such short wavepackets spread over a segment of length $L$ each.
At fixed $L$, that is at a fixed value of 
 the group velocity of each wave packet, the fraction of the probability measure of the centipede 
 that will feel boundary effects over an interval of $L$ units of time 
corresponds to the two short wave packets that are close to the first and last leg of the centipede.
 It is  is proportional to $G^{-1} = L/N$, which goes to zero with $N$.
 At a fixed value of the intensive variable $G/N$ (the maximal density of gaps along the centipede), the proposed concatenation  
 of short wave packets should become a better approximation to a long wave packet 
 traveling at group velocity $2G/N$ when $N$ grows.\\

\begin{figure}
\begin{center}
\includegraphics[width= 16cm,keepaspectratio]{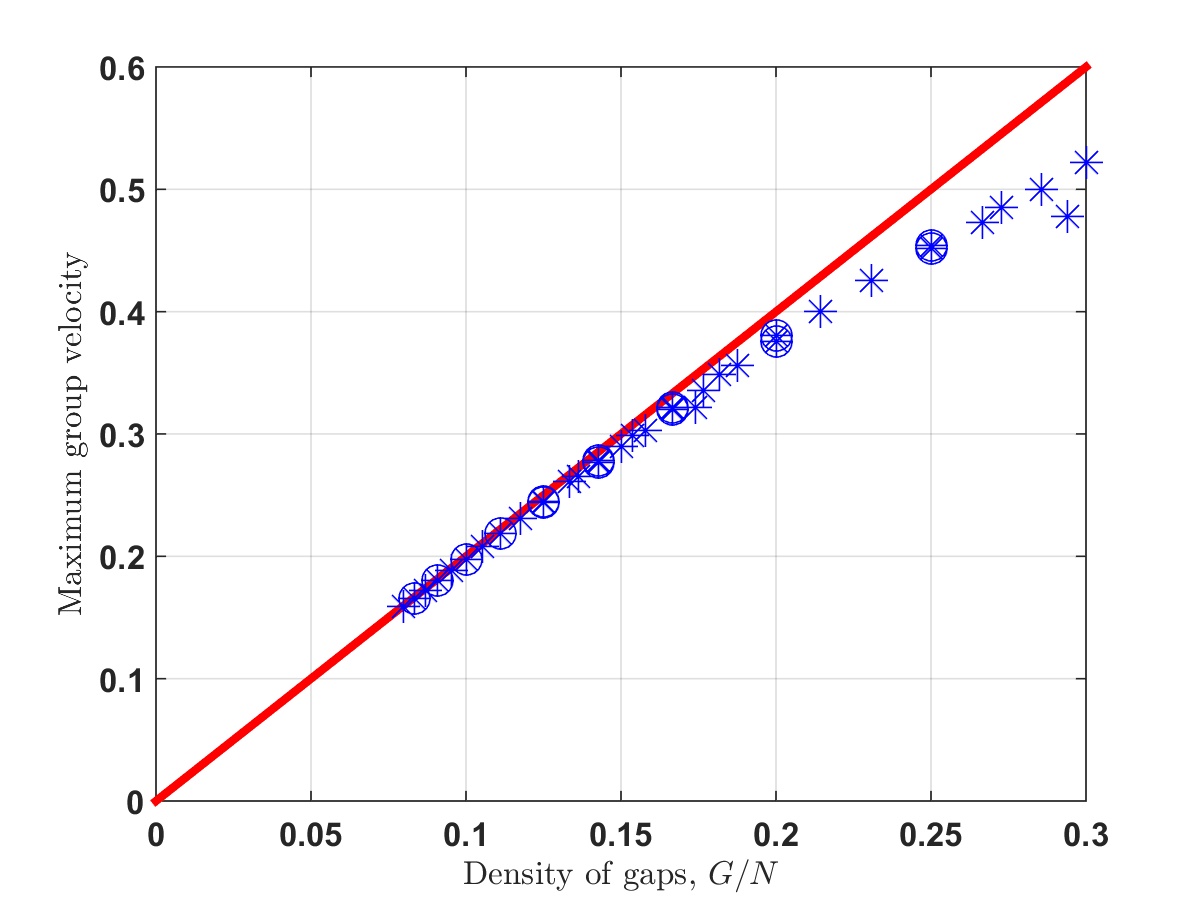}
\end{center}
\caption{The maximal group velocity as a function of the density of gaps, $G/N$, for densities lower than 
$1/3$ (as $3$ is the largest integer for which the linear prediction $2/L$, plotted in red, is larger 
 than the saturation value $V^{\infty}$ worked out in \cite{quantumCentipedes}), and $N$ larger than 10 (so that the saturation value is close to the value $V^{\infty}$). Only values with $G>1$ are shown. There are 14 data points with $G>1$ such that $N/G$
 is an integer (these points are circled). 
The  values of  $N$ range between $10$ and $25$.  The saturation effect  is visible at high densities corresponding to an average 
 length or $L=4$ (the actual value is below the linear prediction by $9-10\%$), $L=5$ (by $5-6\%$), $L=6$  (by $3-4\%$).}
\label{densityCheck}
\end{figure}

 Plotting the maximal group velocities as a function of the 
 maximal allowed density of gaps $G/N$ reveals a linear growth, with negative corrections 
 due to the saturation phenomenon (see Fig. \ref{densityCheck}). The fact that the linearity has been 
 tested numerically within 4 percent of the predicted value at $L=6$ for low values of $N$ 
    suggests lower bounds on the group velocities that can be reached at large $N$ for a fixed value of the 
 denisty of gaps.  For instance a finite group velocity of about $0.26$ should be reachable
 in the  large-$N$ limit with a maximal allowed density of gaps of about 16 percent, or $G = 0.16N$.
  Values of $G$ that are not divisors of $N$ roughly yield a growing maximal group velocity as a function of the
 density, even though the above argument is not valid for them, as the group velocities of the short wave packets 
 cannot be uniform across the concatenation. This effect should become less relevant when $N$ goes to infinity.\\

\subsection{Upper bound on the region of  linear growth of the maximal group velocity from the saturation value}


 On the other hand, at a fixed value of $N$, the maximal group velocity is expected to be a growing 
 function of the total number $G$ of gaps allowed, and to be smaller than the value $V^{(N)}$
 obtained if no global constraint is imposed.
  This value decreases
 exponentially fast as a function of $N$, to reach a limit $V^{(\infty)}$ expressed in integral form in \cite{quantumCentipedes}
 and evaluated numerically to be  $V^{\infty}\simeq 0.570$. The 
  linear growth of the maximal group velocity in the globally constrained model as a function of 
  the density of gaps $G/N$  can therefore 
 not keep going after this value of the group velocity is reached. This yields an upper 
   bound $G_{crit}$ on the number of gaps for which the maximal group velocity
 can be estimated by linear extrapolation of the  model with  strongest global constraint:
\begin{equation}
\frac{2(G_{crit})}{N} = V^{\infty},\;\;\;{\mathrm{ i.e.}}\;\;\; 
G_{crit} \simeq [ 0.2852 N  ].
\end{equation}
 This upper bound grows as $O(N)$ and thanks to the fast convergence 
 to $V^{\infty}$ yields numerically testable cases (some of which are visible on Fig. 
 \ref{linearityCheck}).\\  
 
%

\section{Discussion}

%

 The typical  ballistic behaviour of quantum walkers survives the strongest global constraint 
   imposed on the centipede, but the group velocities of the model go to zero when the number of legs goes
 to infinity. This contrasts with the large-$N$ behaviour of wave fronts of 
 quantum centipedes with local constraints only identified in \cite{quantumCentipedes}. The dimension
 of the Hilbert space of the strongly constrained quantum centipede grows linearly, not exponentially, with $N$.
  In the language of spin models or lattice gas, only one variable is allowed to flip compared to the 
  tightest configuration. 
Intuitively, the flipped variable has to travel all the way through the centipede to contribute to the spreading of a 
 wave front at the two opposite ends of the centipede, which penalizes the group velocity.\\

 At larger values of the allowed number of gaps, the group velocity is observed to grow linearly
 as a function of the number of gaps, so that the global constraint    can yield 
 a strictly positive  fraction  of the saturation value   of the group velocity,
 provided the maximal allowed  density of gaps in the configuration (the intensive variable 
 $(S-N +1)/N$) is strictly positive.  
  The validity of this linear  regime of growth is bounded from above by the saturation value.
   Moreover, when the allowed number of gaps $G$ grows from 1 to $\alpha N$,  wave packets constructed 
 by concatenating $G$ copies of the exacty-solved maximally-constrained  model  (with $L = [N/G]$ legs), each traveling 
at the maximal velocity group $2/L$, are less deformed when $L$ is larger. The value $L=3$
 is critical in the sense that the first detectable move of a wave front centered on a segment of a  centipede with three
 legs  becomes adjacent to the next segment in the concatenation. The corresponding  value of the 
 group velocity, at $2/3$, is larger than the exact value of $V^{\infty}$, and $L=3$ is the lowest value of $L$
  with this property. 
It would be interesting to obtain exact expressions for the 
 dispersion relation of the Hamiltonian of Fourier modes at a fixed value of $G$, which is beyond the scope
 of the present paper, and to make the concatenation argument rigorous in terms of 
 interactions in spin models. The description
 of the Hilbert space in terms of compositions of the number of legs could be relevant 
 for that purpose, and shed some light on the interpolation between the linear regime of growth and the saturation 
 regime, as a function of the density of gaps.

\section*{Acknowledgments}
 I would like to thank Sakura Sch\"afer-Nameki and Julien Guyon for correspondence. This work 
 was supported by the Research Development Fund of Xi'an Jiaotong-Liverpool University (RDF-14-01-34).

\end{document}